\documentclass[a4paper,12pt]{article}

\pdfoutput=1

\usepackage{amssymb}
\usepackage{cite}
\usepackage{epsf}
\usepackage{epsfig}
\usepackage{subfigure}
\usepackage{mathtools}
\usepackage{hhline}
\usepackage{float}
\usepackage{multirow}
\usepackage{nicefrac}
\usepackage{epstopdf}
\usepackage{slashed}
\usepackage{xcolor}
\usepackage{url}
\usepackage{titlesec}
\usepackage{ulem}
\usepackage{hyperref}

\usepackage{afterpage}

\graphicspath{{figures/}} 

\newcommand{\GeV}{\mathrm{GeV}}
\newcommand{\TeV}{\mathrm{TeV}}

\newcommand{\eps}{\epsilon}

\newcommand{\Lcal}{\mathcal{L}}

\newcommand{\ol}[1]{\overline{#1}}

\newcommand{\br}[2]{\mathrm{BR} \left({#1}\to{#2}\right)}

\newcommand{\met}{E_T^\mathrm{miss}}

\usepackage{amsmath}	

\begin{document}

\begin{titlepage}

\begin{flushright}
 {\tt
CTPU-PTC-23-38  \\
}
\end{flushright}

\vspace{1.2cm}
\begin{center}
{\Large
{\bf
Current status on pair-produced muon-philic vectorlike leptons 
in multilepton channels at the LHC \\
}
}
\vskip 2cm
Junichiro Kawamura$^{\ a}$~\footnote{junkmura13@gmail.com}, 
Seodong Shin$^{\ a, b}$~\footnote{sshin@jbnu.ac.kr}    

\vskip 0.5cm

{\it $^a$
Particle Theory  and Cosmology Group, Center for Theoretical Physics
of the Universe, Institute for Basic Science (IBS), Daejeon, 34126, Korea
}\\[3pt]

{\it $^b$
Laboratory for Symmetry and Structure of the Universe, Department of Physics, Jeonbuk National University, Jeonju, Jeonbuk 54896, Korea}\\[3pt]

\vskip 1.5cm

\begin{abstract}
In this work, we obtain the current limits on the pair production of vectorlike leptons 
decaying to a Standard Model gauge boson and a lepton in the second generation
using the Run-2 data at the LHC.
Since there is no dedicated search out of Run-2 data, 
we recast the ATLAS analyses searching for 
the type-III seesaw heavy leptons in the multi-lepton channels.
There is no limit for the $SU(2)_L$ singlet vectorlike lepton beyond about 100 GeV, 
while the limit is about 780 GeV for the doublet one. 
Thus, dedicated searches for the vectorlike leptons are necessary, 
especially for the singlet one.  
We also study the general cases of the vectorlike lepton decays 
and future sensitivities at the HL-LHC.
\end{abstract}
\end{center}
\end{titlepage}

\clearpage

\setcounter{footnote}{0}
\section{Introduction}

Vectorlike (VL) fermions are key ingredients in many new physics models 
beyond the Standard Model (SM) adopted 
to resolve both theoretical and experimental issues. 
Since chiral fermions in the fourth family are excluded experimentally~\cite{Djouadi:2012ae,Eberhardt:2012gv}, 
these are considered to be vectorlike and their masses are given independently to the Higgs mechanism in the SM. 
The VL fermions are introduced in, for instance, 
supersymmetric models~\cite{Martin:2009bg,Graham:2009gy,Moroi:2011aa,Endo:2011xq,Endo:2011mc,Dermisek:2017ihj,Dermisek:2018hxq,Araz:2018uyi}, 
gauge mediated supersymmetry breaking scenario~\cite{Dine:1981gu,Alvarez-Gaume:1981abe,Nappi:1982hm,Dine:1993yw,Dine:1994vc,Dine:1995ag}, 
composite Higgs models~\cite{Panico:2015jxa,Cacciapaglia:2020kgq,Cacciapaglia:2022zwt}, KSVZ axion models~\cite{Kim:1979if,Shifman:1979if}, 
axion-like particle models~\cite{Kim:2008hd,Marsh:2015xka,DiLuzio:2020wdo,Choi:2020rgn}, 
alternative solutions of the strong CP problem~\cite{Mohapatra:1978fy,Babu:1989rb}, 
two Higgs doublet model augmented by VL fermions~\cite{Dermisek:2015hue,Dermisek:2015oja,Dermisek:2015vra,Dermisek:2016via,Dermisek:2019vkc,Dermisek:2019heo,Dermisek:2020gbr,Dermisek:2022kyh,Dermisek:2022xal,Dermisek:2022aec,Dermisek:2023tgq}, 
and models for gauge coupling unification ~\cite{Dermisek:2012as,Dermisek:2012ke}.

Among the fourth family fermions, 
VL leptons (VLLs) with nonzero lepton number 
play unique roles in constructing lepton-philic dark matter (DM) models~\cite{Bai:2014osa,Chang:2014tea,Kawamura:2017ecz,Calibbi:2018rzv,Kawamura:2020qxo,Kawamura:2022uft}, 
mirror sector models~\cite{Kawamura:2018kut,Dunsky:2019api,Jung:2019fsp}, 
and explanations for the muon anomalies~\cite{Kannike:2011ng,Dermisek:2014cia,Poh:2017tfo,Raby:2017igl,Arnan:2019uhr,Kawamura:2019rth,Kawamura:2019hxp,Chun:2020uzw,Frank:2020smf,Crivellin:2020ebi,Dermisek:2020cod,Dermisek:2021ajd,Lee:2021gnw,Lee:2022nqz,Kawamura:2022fhm,Abdallah:2023pbl}~\footnote{
The latest experimental result~\cite{Muong-2:2023cdq} 
confirms the previous results~\cite{Bennett:2006fi,Abi:2021gix} 
which might deviate from the SM prediction~\cite{Aoyama:2020ynm}.
}. 
Interestingly, the lightest VLL is expected to be in the reach 
of the Large Hadron Collider (LHC) or High-Luminosity (HL)-LHC.
Both ATLAS and CMS collaborations search for the 
pair production of VLLs each 
of which dominantly decays to a SM boson and a tau lepton~\cite{CMS:2019hsm,CMS:2022nty,ATLAS:2023sbu}. 
For the doublet VLL~\footnote{Throughout this work, 
doublet (singlet) for VLL means iso-doublet (iso-singlet)
under the $SU(2)_L$ gauge symmetry.}, 
the ATLAS search excludes the mass range of 
$130< m_{\mathrm{VLL}}<900~\GeV$~\cite{ATLAS:2023sbu}, 
and the CMS search excludes the mass up to 1045 GeV~\cite{CMS:2022nty}.
The singlet VLL is less constrained and the limit is less than $150~\GeV$~\cite{CMS:2022nty}.
Prospects of such VLLs at the future colliders are discussed in Ref.~\cite{Bhattiprolu:2019vdu}.
The pair productions of the VLLs decaying to a SM boson and a muon (neutrino) are studied by the ATLAS~\cite{ATLAS:2015qoy} and the theorists~\cite{Dermisek:2014qca,Kumar:2015tna} using the Run-I data. 
The limits are obtained for $114 \le m_{\rm VLL} \le 176~\GeV$ in the case of the singlet VLL 
and $m_{\rm VLL} \lesssim 500$ GeV when the neutral component of the lightest doublet VLL 
dominantly decays to a $W$ boson and a muon~\cite{Dermisek:2014qca}.
There are also studies for the VLL produced from cascade decays 
of extra neutral Higgs bosons~\cite{Dermisek:2015hue,Dermisek:2015oja,Dermisek:2015vra,Dermisek:2016via,Dermisek:2022kyh,Dermisek:2022xal}, 
and signals from the VLLs decaying to a DM particle~\cite{Guedes:2021oqx} 
or $Z^\prime$ boson~\cite{Kawamura:2021ygg}.

In this paper, we study pair-productions of the VLLs, through the Drell-Yan process,
decaying to the second generation lepton, namely {\it muon-philic} VLL.
Such VLL is well motivated to explain the experimental anomalies 
in the muon $g-2$~\cite{Muong-2:2004fok,Aoyama:2020ynm,Muong-2:2021ojo} 
and the semi-leptonic $B$ decays~\cite{LHCb:2017avl}.~\footnote{
The LHCb has recently announced the new result of $R_{K^{(*)}}$ 
consistent with the SM expectation~\cite{LHCb:2022zom} 
which requires efforts such as separate measurements of the branching ratio 
at Belle-II~\cite{Belle-II:2022fky}.}
In this work, we obtain the current limits using the Run-2 data at $\sqrt{s} = 13$ TeV 
by simply recasting the ATLAS analyses searching for the triplet lepton 
in the type-III seesaw~\cite{ATLAS:2020wop,ATLAS:2022yhd}.
We then estimate expected sensitivities at the HL-LHC using the same channels.

This paper is organized as follows. 
We briefly explain the VLLs in Sec.~\ref{sec-VLL} 
and discuss the analysis strategy in Sec.~\ref{sec-anal}.
Our main results are shown in Sec.~\ref{sec-rslt}.
Finally, we summarize the paper in Sec.~\ref{sec-concl}.

\section{Vectorlike lepton scenario}
\label{sec-VLL}

\begin{figure}[t]
 \centering
\includegraphics[height=0.48\hsize]{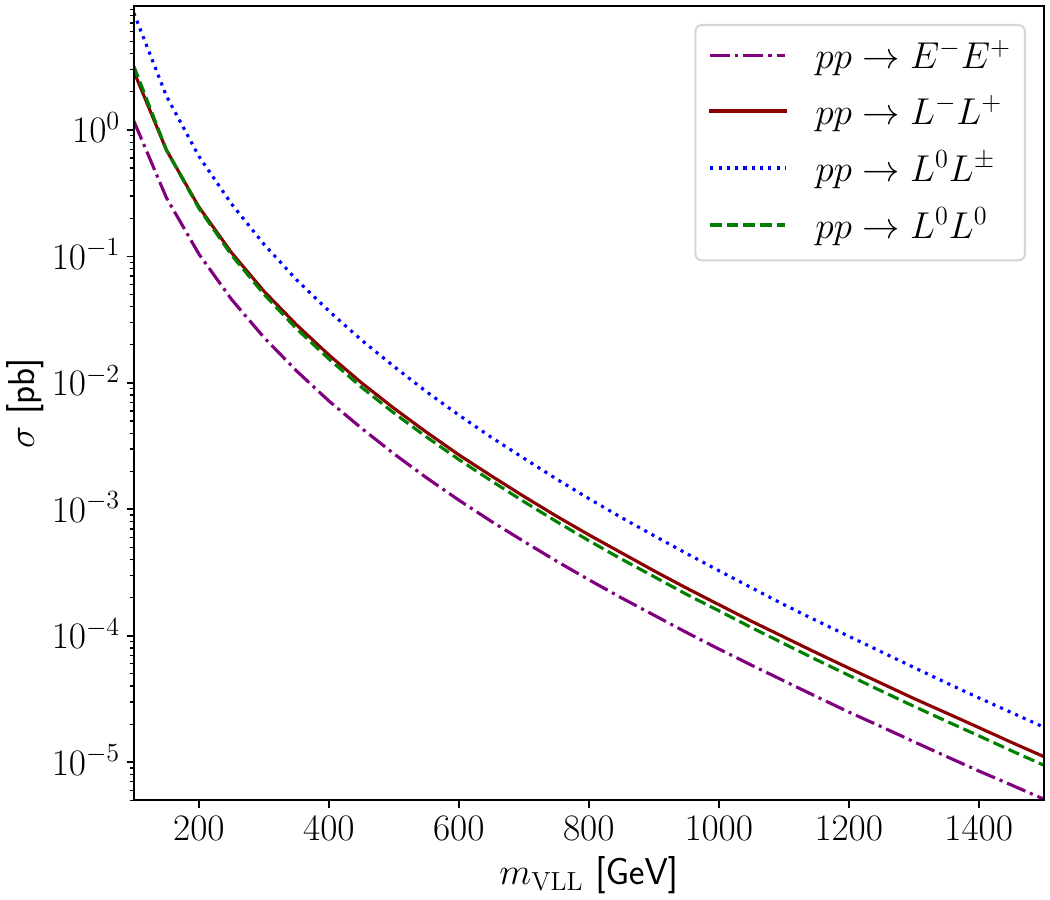}
\caption{\label{fig-prodNLO}
The NLO production cross sections at $\sqrt{s}=13$ TeV.
} 
\end{figure}

In this work, we consider the two kinds of VLLs 
whose gauge quantum numbers are given by 
\begin{align}
 (E, \ol{E}) = (\mathbf{1}_{-1}  , \mathbf{1}_{1}), \quad 
 (L, \ol{L}) = (\mathbf{2}_{-1/2}, \mathbf{2}_{1/2}),    
\end{align}
where $\mathbf{r}_Y$ is a representation $\mathbf{r}$ under $SU(2)_L$ 
and the index $Y$ is the hypercharge under $U(1)_Y$. 
Here the electric charge is given by $Q = T_3 + Y$, with $T_3$ being isospin.  
The doublet $L$ has two components denoted by $(L^0, L^-)$, 
where the superscript is the electric charge. 
These VLLs can be considered as the fourth 
generation leptons since $L$ and $E$ have the same quantum numbers as the SM leptons.

For simplicity, we assume that one of the VLLs is heavy enough 
so that the lightest charged VLL is almost either singlet or doublet~\footnote{
See, e.g. Ref.~\cite{Dermisek:2013gta} for discussions about the mass mixing 
induced by the electroweak (EW) symmetry breaking for $m_E \sim m_L$.
}. 
Under this assumption, 
the mass of the neutral VL lepton $L^0$ is the same as 
that of the charged counterpart $L^-$ approximately. 
We further assume that 
the VLLs exclusively couple to the second generation of the SM leptons 
to avoid lepton flavor violations strongly constrained by the experiments. 
In Fig.~\ref{fig-prodNLO}, 
we show the next-to-leading order (NLO) cross sections 
of the VLL pair production, $pp \to E \ol{E}$ or $L \ol{L}$, 
at $\sqrt{s} = 13$ TeV. 
The cross sections at the fixed-order NLO 
are calculated by using \texttt{MadGraph5}~\cite{Alwall:2014hca}
based on the \texttt{UFO} model file~\cite{Degrande:2011ua} 
developed in Ref.~\cite{Ajjath:2023ugn}.

The lightest charged VLL decays to a SM boson and a lepton in the second generation 
$E^-/L^- \to W^- \nu_\mu, Z \mu^-, h \mu^-$ 
if there is no new particle lighter than the VLL.
The neutral one $L^0$ decays as $L^0 \to W^+ \mu^-$~\footnote{
The decays $L^0 \to Z \nu_\mu$, $h\nu_\mu$ will be open 
if we introduce additional SM singlet VLL, 
but we do not consider these modes for simplicity. 
}.
If the mass of the lightest charged VLL is large enough compared
to the SM boson masses, the branching ratios are given by  
\begin{align}
 \br{E^-}{W^- \nu_\mu}:\br{E^-}{Z \mu^-}:\br{E^-}{h \mu^-} =&\ 2:1:1\,, 
\label{eq-BrEmin}
\\
 \br{L^-}{W^- \nu_\mu}:\br{L^-}{Z \mu^-}:\br{L^-}{h \mu^-} =&\ 0:1:1\,.  
\label{eq-BrLmin}
\end{align}
as expected from the Goldstone boson equivalence theorem. 
Nevertheless, we shall allow other values of the branching ratios 
since the above patterns can be broken  
when some of the new Yukawa couplings have a certain hierarchical structure~\cite{Dermisek:2014qca,Dermisek:2019vkc}.

\section{Analysis} 
\label{sec-anal}

In this section, 
we summarize our strategy to recast the ATLAS analyses searching  
for the triplet leptons in the type-III seesaw model~\cite{ATLAS:2020wop,ATLAS:2022yhd}. 
In Ref.~\cite{ATLAS:2020wop}, 
the signals are composed of two light leptons, 
i.e. $e$ or $\mu$, with at least two jets and large missing energy, 
where the two jets are from a $Z/W$ boson decay. 
Both opposite sign (OS) and same sign (SS) leptons are searched in the analysis,  
but in our case the signal will not contribute to the signal regions (SRs)
with the SS leptons because our neutral VLL, $L^0$, 
is a Dirac but not a Majorana fermion.
We also note that the SRs containing muons will be the most important 
to identify the muon-philic VLL.
In Ref.~\cite{ATLAS:2022yhd}, the signals with 3 or 4 leptons are studied. 
Overall, our final states should be either di-leptons 
with a hadronic $Z/W$ boson or 3/4 charged leptons involving muons.

We simulate the signal events by using \texttt{MadGraph5}~\cite{Alwall:2014hca} 
based on the \texttt{UFO} model file generated by \texttt{FeynRules}~\cite{Alloul:2013bka,Christensen:2008py}.  
The decays of the VLLs are handled by~\texttt{MadSpin}~\cite{Artoisenet:2012st}, 
and then the parton-level events are 
showered and hadronized by \texttt{PYTHIA8}~\cite{Sjostrand:2007gs}.     
The pair productions of the VLLs are simulated up to one additional parton using the MLM matching~\cite{Caravaglios:1998yr} 
with $\texttt{xqcut} = m_{\mathrm{VLL}}/10$. 
The fast detector simulation is performed 
by \texttt{Delphes3.4.2}~\cite{deFavereau:2013fsa} with the default ATLAS card. Jets are reconstructed  
using the anti-$k_T$ algorithm~\cite{Cacciari:2008gp,Cacciari:2011ma} 
with $\Delta R = 0.4$. 
The same cut conditions are applied for the generated events 
as the ATLAS analyses~\cite{ATLAS:2020wop,ATLAS:2022yhd}, 
but the object-based missing transverse momentum significance $S(\met)$ is 
calculated by the approximated one $S(\met) := \met/\sqrt{H_T}$, 
where $\met$ is the missing transverse momentum 
and $H_T$ is the sum of transverse momenta 
of visible particles~\cite{ATLAS-CONF-2018-038}. 
Note that the Higgs decays are irrelevant for the SRs 
where the b-tagged jets are vetoed.~\footnote{
The b-veto is not required in the $Q2$ SR
where the total charge of the leptons are $\pm 2$, 
but the rate is tiny in the SR because the VLLs are Dirac fermions. 
}

We shall study the pair productions of the VLLs: 
\begin{align}
 pp \to E^+ E^-, L^+ L^-, L^\pm L^0, L^0 L^0,  
\end{align}
followed by the decays of the VLLs to a lepton in the second generation 
and a $Z$ or $W$ boson. 
We label the processes as $XY$-$VV^\prime$,  
where $X,Y$ are pair-produced VLLs. 
In this labeling, $X,Y = F^\pm,L^0$, 
where $F^\pm = E^\pm, L^\pm$.
The kinetic distributions will be common for $E^\pm$ and $L^\pm$, 
while the production cross sections are different. 
$V~(V^\prime)$ is a SM EW gauge boson from the decay of $X$ ($Y$). 
The processes labeled by $F^+F^-$-$WW$ will not contribute to the SRs 
because the leptons produced from the VLL decays 
are neutrinos, and hence the numbers of leptons are not enough to pass the cuts.
Altogether, the relevant processes are 
$F^+F^-$-$ZZ$, $F^+F^-$-$WZ$, $L^0F^-$-$WW$, $L^0F^-$-$WZ$ 
and $L^0L^0$-$WW$,  
so we simulate these 5 processes.

We use the statistic variable $q_\mu$ defined as~\cite{Cowan:2010js},  
\begin{align}
\label{eq-defqmu}
 q_\mu = -2 \log \frac{L(\mu, \hat{\hat{b}})}{L(\hat{\mu}, \hat{b})},  
\end{align}
where the likelihood function is given by 
\begin{align}
 L(\mu, b) := \prod_{i} \frac{\left(\mu s_i + b_i\right)^{n_i}}{n_i !} e^{-(\mu s_i + b_i)}
              \times \frac{1}{\sqrt{2\pi\sigma_i^2}} \mathrm{exp}
              \left[ -\frac{(b_i-b_i^0)^2}{2\sigma_i^2}\right].   
\end{align}
Here, $n_i$, $s_i$, $b_i^0$ and $\sigma_i$ 
are respectively the number of observed, signal, background events
and the error of the number of backgrounds in the bin labeled by $i$.  
Note that the SRs in the analysis are exclusive with each other 
because these have the different number of leptons.  
In Eq.~\eqref{eq-defqmu}, 
$\hat{\hat{b}}$ is the number of backgrounds 
which maximizes $L$ for a given $\mu$, 
and $(\hat{\mu}, \hat{b})$ are the values of $(\mu,b)$ which maximizes $L$. 
We only consider the uncertainties from the backgrounds 
which are typically the dominant sources for the errors. 
The 95\% C.L. limit is defined as~\cite{CMS-NOTE-2011-005}
\begin{align}
CL_s := \left.
\frac{1-\Phi\left(\sqrt{q_{\mu}}\right)}{\Phi\left(\sqrt{q_\mu^A} -\sqrt{q_\mu}\right)}
 \right|_{\mu=1} < 0.05,  
\end{align}
where $\Phi$ is the cumulative distribution function of the normal distribution. 
Here, $q^A_\mu$ is the test statistics in Eq.~\eqref{eq-defqmu} 
by replacing $n_i \to b^0_i$. 
For the future limits, the significance is given by~\cite{Cowan:2010js} 
\begin{align}
 Z = \sqrt{q_\mu|_{\mu=1, n=b}}.   
\end{align}
The number of signal and background events (its error) are rescaled as $R_{\Lcal}$ ($\sqrt{R_\Lcal}$), 
where  $R_{\Lcal}$ is the ratio of the integrated luminosity $\Lcal$. 
The number of observed events 
and the estimated backgrounds are shown 
in the ATLAS papers~\cite{ATLAS:2020wop,ATLAS:2022yhd}. 
The number of signal events is calculated as 
\begin{align}
\label{eq-defsi}
 s_i = \Lcal 
         \sum_{P} 
         \sum_{D} 
         \sigma_P \times \mathrm{Br}_D \times \eps^{P,D}_i,  
\end{align}
where $P = F^-F^+, L^0F^\pm, L^0L^0$ 
labels the production of the VLL pairs
whose cross section is $\sigma_P$, 
and $D = ZZ, WW, WZ$ labels the decays whose branching fraction is 
$\mathrm{Br}_D$ given by $\br{X}{V \ell_2}$ $\times$ $\br{Y}{V^\prime \ell_2}$,
where $\ell_2 = \mu, \nu_\mu$.
$\eps^{P,D}_i$ is the number of events passing the cut in the bin $i$, 
per the number of events generated.
We obtain the values of $\eps^{P,D}_i$ by the simulation. 
For the future limits at the HL-LHC, 
we assume pp collisions at $\sqrt{s} = 13~\TeV$, rather than $13.6$ or $14$ TeV, 
to be conservative and the integrated luminosity of $3~\mathrm{ab}^{-1}$ data.

\section{Results}
\label{sec-rslt}

\begin{figure}[t]
 \centering
\includegraphics[width=0.95\hsize]{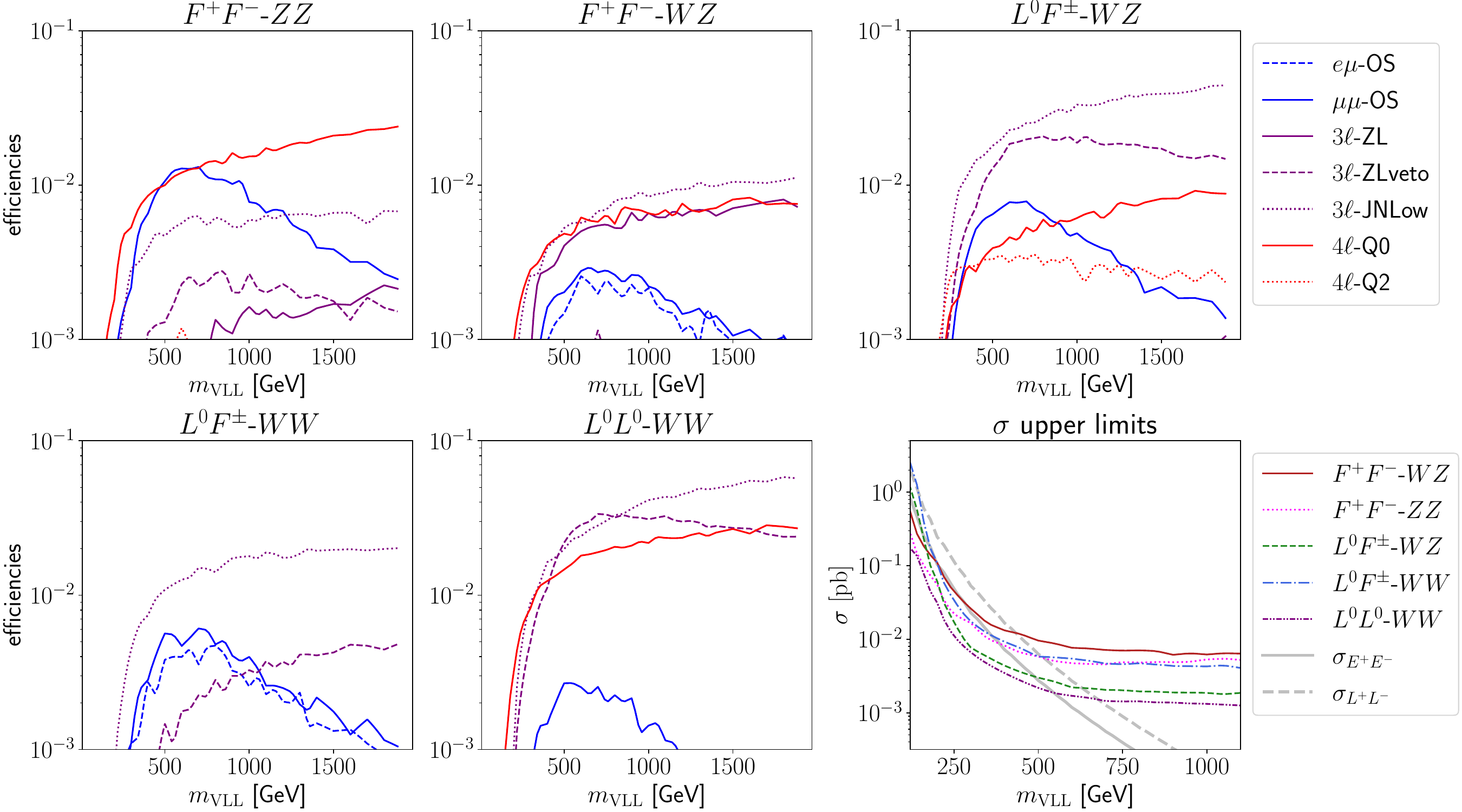}
\caption{\label{fig-effs}
Efficiencies from the processes studied in this work. 
Here, $F^\pm = E^\pm, L^\pm$ is one of the charged VLLs.
In addition to the efficiencies, we show the $95\%$ C.L. upper bounds on the production cross sections 
based on the individual SRs on the bottom-right panel. 
The gray lines show the production cross sections of the VLL pairs. 
} 
\end{figure}

In this section, we explain our analysis results on the signal efficiencies, the current limits, and the future limits after the full running of the HL-LHC.
Figure~\ref{fig-effs}, except the bottom-right panel, 
shows the efficiencies of the processes in the relevant SRs: 
$e\mu$-OS and $\mu \mu$-OS for the two lepton final states 
defined in Table 2 of Ref.~\cite{ATLAS:2020wop}, 
and $3\ell$-ZL, $3\ell$-ZLveto, $3\ell$-JNLow, 
$4\ell$-Q0 and $4\ell$-Q2, defined in Tables 2 and 3 of Ref.~\cite{ATLAS:2022yhd}.
Here, the label ZL represents leptonically decaying $Z$ boson, 
and it is required to exist in $3\ell$-ZL and $3\ell$-JNLow, 
while it is vetoed in $3\ell$-ZLveto. 
The jet multiplicity is required to be less than one in $3\ell$-JNLow. 
For the $4\ell$ SRs, 
$Q$ is the absolute value of the sum of the lepton charges in final states 
and it is required to be zero or two. 
Note that the $4\ell$-Q2 SR is possible only when some of the leptons are missing and hence the corresponding efficiencies are lower than the others.
The efficiencies are smaller than $10^{-3}$ in the other SRs, 
namely $ee$-OS and $\ell\ell^\prime$-SS ($\ell, \ell^\prime = e,\mu$), 
because the VLLs decay to the second generation leptons and they preserve the lepton numbers. 
Note that the efficiencies in the $2\ell$ SRs decrease for the heavier VLL mass 
since the event selection requirement $60 < m_{jj}< 100$ GeV in Ref.~\cite{ATLAS:2020wop} is less likely to be satisfied. 
Whereas, those in the SRs with $3$ or $4$ leptons increase for heavier VLL due to larger lepton transverse momentum. 
All of the efficiencies drop for light VLL where $m_{\mathrm{VLL}} \lesssim 200$ GeV, 
due to the requirement of two leptons with $p_T>40\GeV$. 
We point out that the $4\ell$-Q0 SR and $\mu \mu$-OS are the dominant event selection category for the $F^+F^-$-$ZZ$ channel where the even numbers of leptons are predicted. 
For the channels involving a $W$ boson, 
the $3\ell$-JNLow SR is the most important, 
particularly for heavier VLL with $m_{\rm VLL} \gtrsim 700$ GeV. 
There are three leptons if a $W$ boson decays leptonically with about 20\% BR, 
and the cuts for the invariant masses of 2-4 leptons systems and $H_T + \met$ 
are not applied in the $3\ell$-JNLow. 
The tighter cut for $m_T(\ell_{1,2})$ than the other $3\ell$-SRs 
are easily satisfied for heavier VLLs, 
and thus the efficiency increases for heavier VLLs in this SR. 

In addition to the efficiencies, 
we show the $95\%$ C.L. upper bounds on the production cross sections 
based on the individual SRs on the bottom-right panel 
of Fig.~\ref{fig-effs}. 
The total cross sections of the charged VLL pair-productions 
are shown for comparison. 
The limits can be obtained by multiplying the factor 
from the branching fractions to the SM gauge bosons, 
i.e. $\mathrm{Br}_D$ in Eq.~\eqref{eq-defsi}. 
Note that the limits are from the individual processes, 
and the actual limits on the VLLs shown in the following
will be obtained by combining these processes.

\begin{figure}[t]
 \centering
\begin{minipage}[t]{0.3\hsize}
\centering
\includegraphics[width=0.95\hsize]{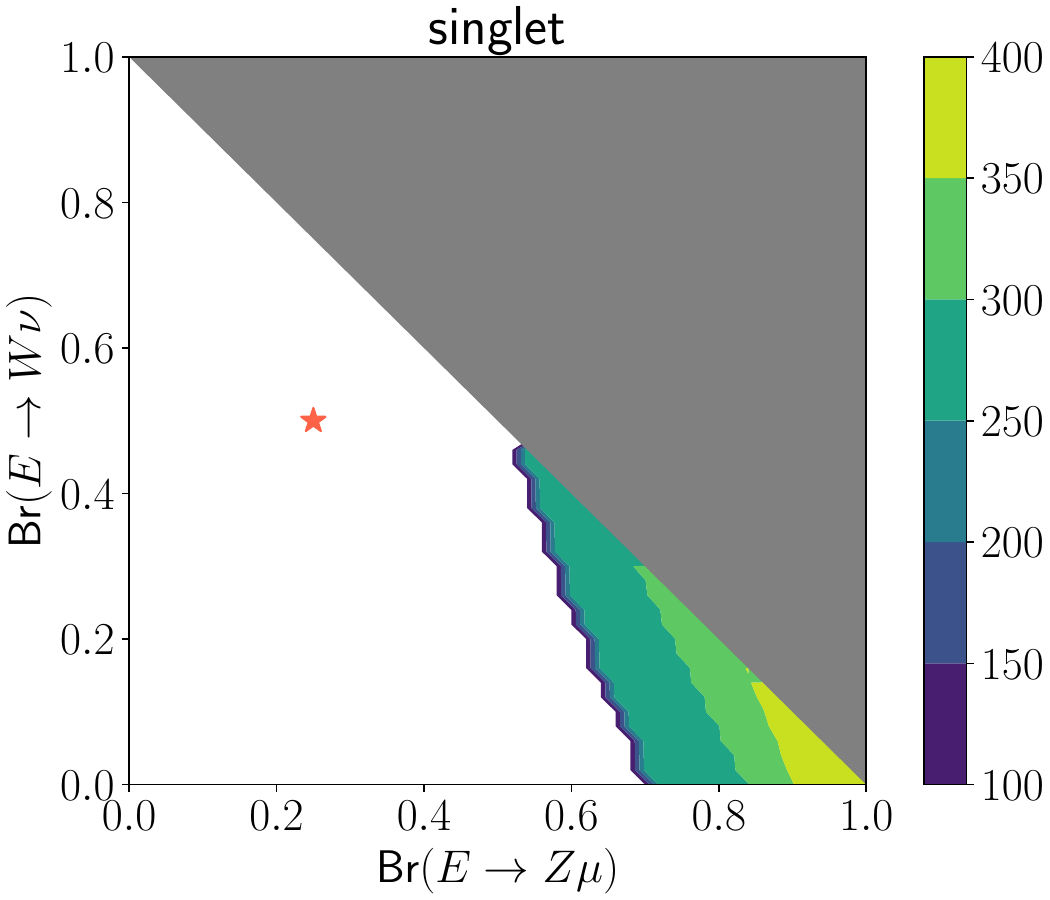} 
\end{minipage}
\begin{minipage}[t]{0.3\hsize}
\centering
\includegraphics[width=0.95\hsize]{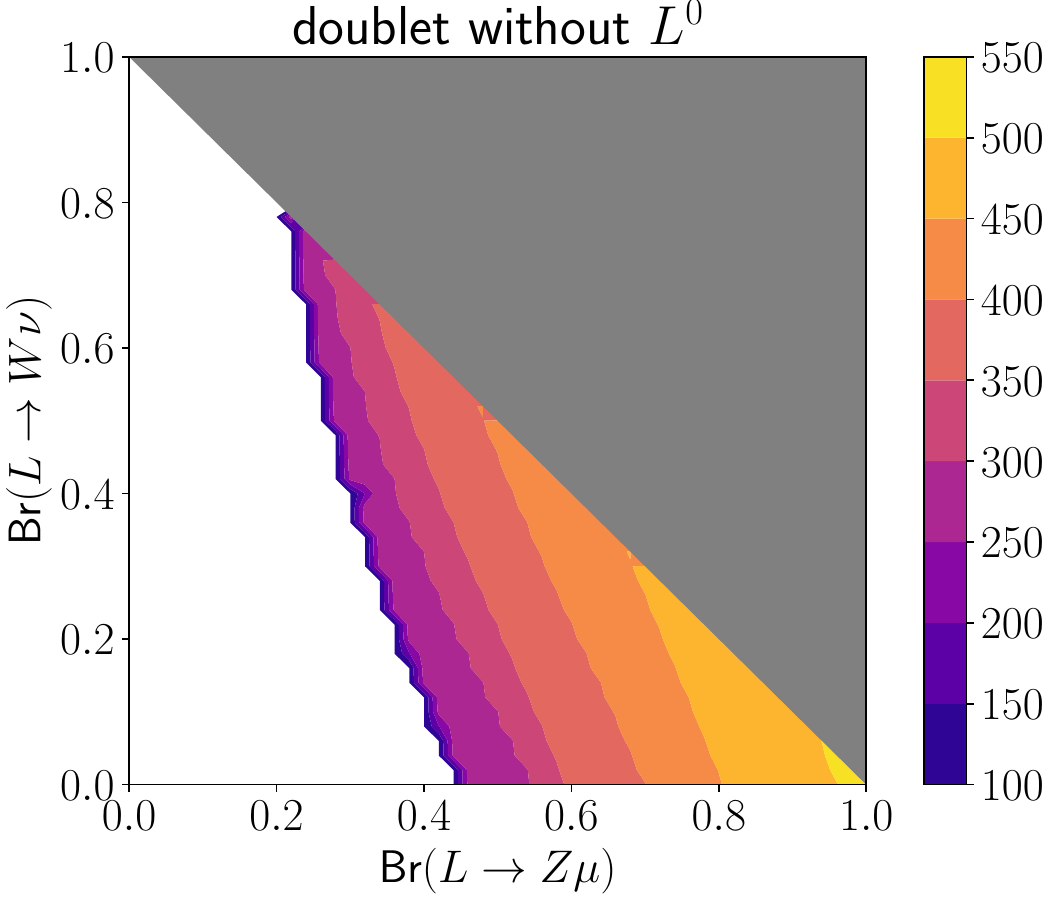} 
\end{minipage}
\begin{minipage}[t]{0.3\hsize}
\centering
\includegraphics[width=0.95\hsize]{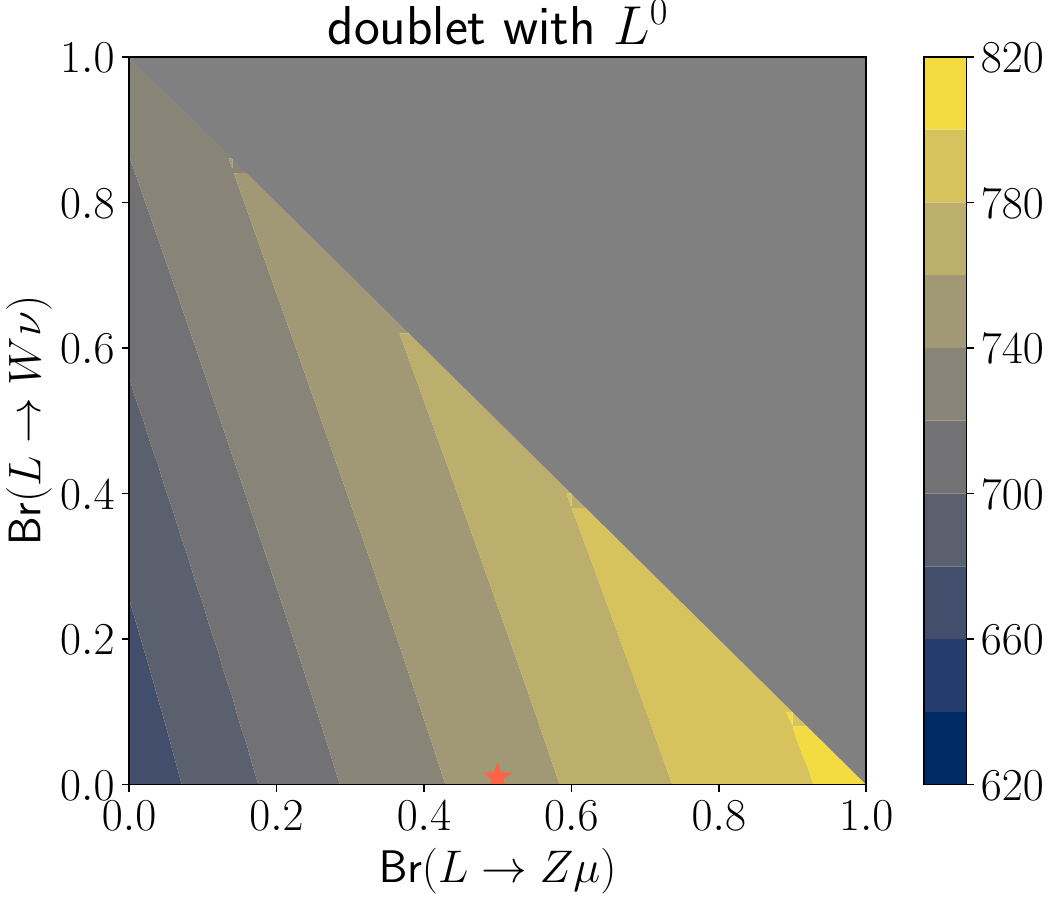}  
\end{minipage}
\caption{\label{fig-lims}
Current $95\%$ C.L. limits on the VLL masses in GeV for the singlet VLL (left), 
doublet without $L^0$ (middle) and doublet with $L^0$ (right). 
There is no limits on the mass in the white region where the corresponding BRs 
are too small. The orange markers indicate the predictions in the minimal scenarios. 
} 
\end{figure}

We obtain the current limits on the VLLs with the signal efficiencies. 
Figure~\ref{fig-lims} shows the 95\% C.L. lower limits on the VLL mass  
in the two dimensional plane of the branching ratios into $W$ boson and $Z$ boson.
The VLL is a $SU(2)_L$ singlet (doublet) in the left (middle and right) panel. 
In the middle panel, only the charged VLL production is taken into account, 
which can be applied to the scenarios with large mass splitting 
between the neutral and charged VLLs, $m_{L_0} \gg m_{L^\pm}$, 
in a theoretical set-up such as mixing with additional neutral particles~\cite{Dermisek:2015oja}.
If the masses of charged and neutral leptons are sufficiently degenerate, 
the results on the right panel are applied. %

In the left panel of Fig.~\ref{fig-lims}, 
we see that the current limit on the singlet VLL mass reaches about 400 GeV 
for large BR($E \to Z \mu) \sim 1$ due to the leptonic decay modes of $Z$ boson.
However, there is no limit for BR($E \to Z \mu) \lesssim 0.5$ due to the small production cross sections of $pp \to EE$. 
In particular, no limit beyond about 100 GeV  exists 
in the case of the minimal singlet VLL model shown in Eq.~\eqref{eq-BrEmin}, 
pointed by the orange marker.
On the other hand, the limit is stronger for the doublet case because of the larger cross sections. 
Without the production of the neutral VLL the limit is 
about $m_{\rm VLL} \gtrsim 500$ GeV 
for $\br{L}{Z\mu} \sim 1$ and $m_{\rm VLL} \gtrsim 150$ GeV
up to $\br{L}{Z\mu} \gtrsim 0.3$. 
With the production of the neutral VLL, 
we obtain the lower limit of VLL mass at least $620$ GeV, 
dominantly determined by $L^0L^0$-$WW$.
The limit reaches to 800 GeV for $\br{L}{Z\mu} \sim 1$, 
and the limit for the minimal doublet VLL scenario in Eq.~\eqref{eq-BrLmin} 
is about 750 GeV.  
Note that the sensitivities are mostly determined by the final states belonging to $3\ell$-ZNLow or $4\ell$-Q0 SRs as seen in Fig.~\ref{fig-effs}, 
requiring high lepton multiplicity.
Therefore, the contours in Fig.~\ref{fig-lims} depend more on BR($E/L \to Z \mu$).

\begin{figure}[t]
 \centering
\begin{minipage}[t]{0.3\hsize}
\centering
\includegraphics[width=0.95\hsize]{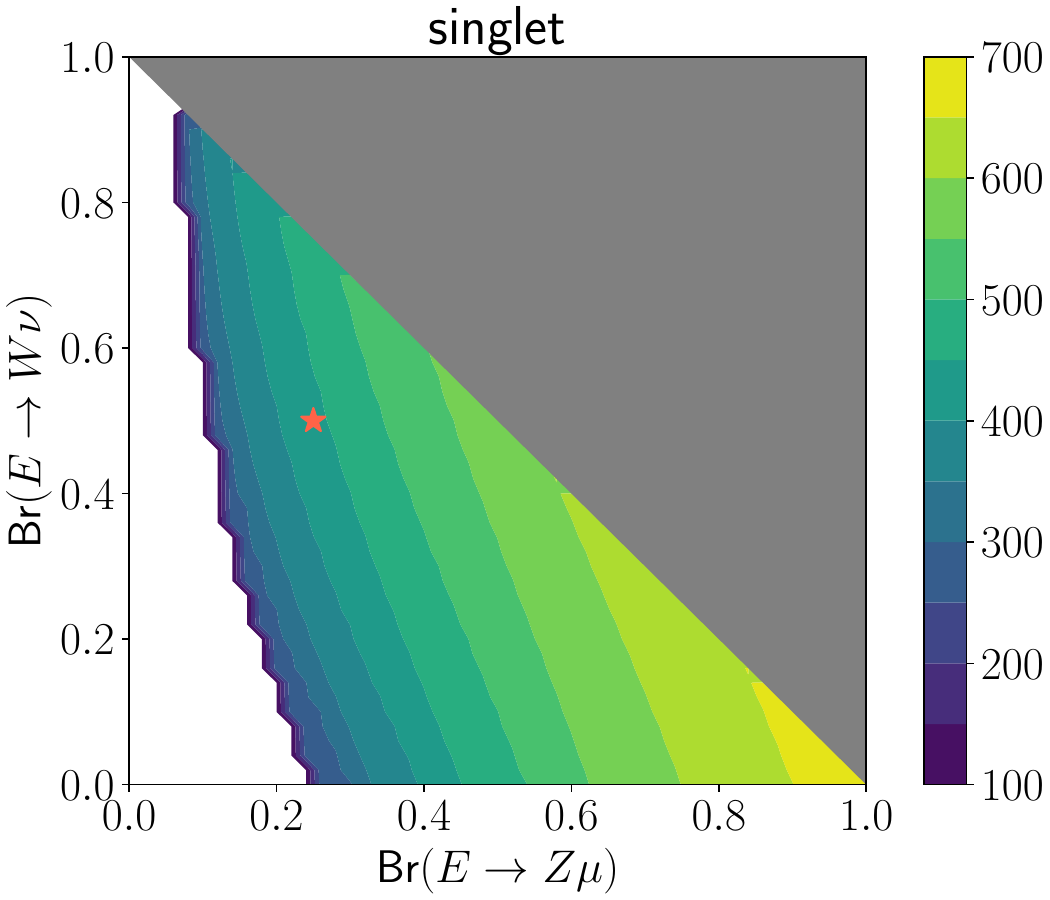} 
\end{minipage}
\begin{minipage}[t]{0.3\hsize}
\centering
\includegraphics[width=0.95\hsize]{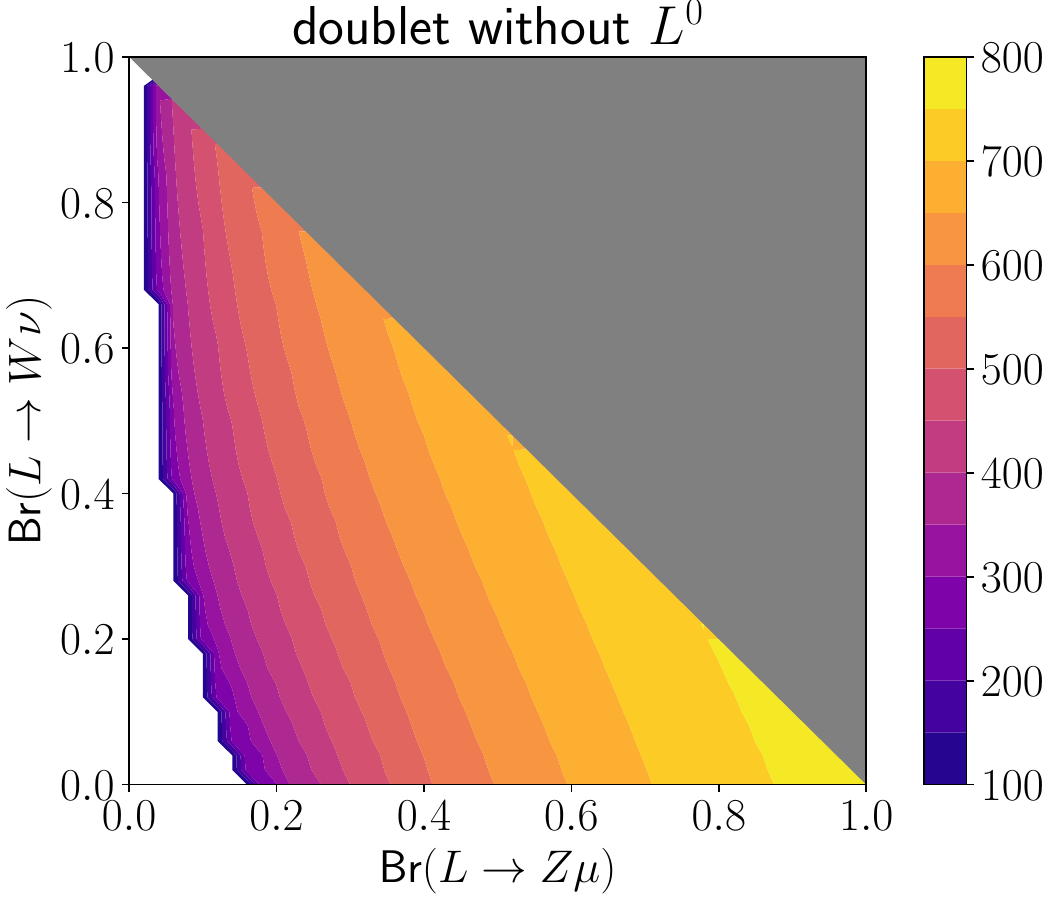} 
\end{minipage}
\begin{minipage}[t]{0.3\hsize}
\centering
\includegraphics[width=0.95\hsize]{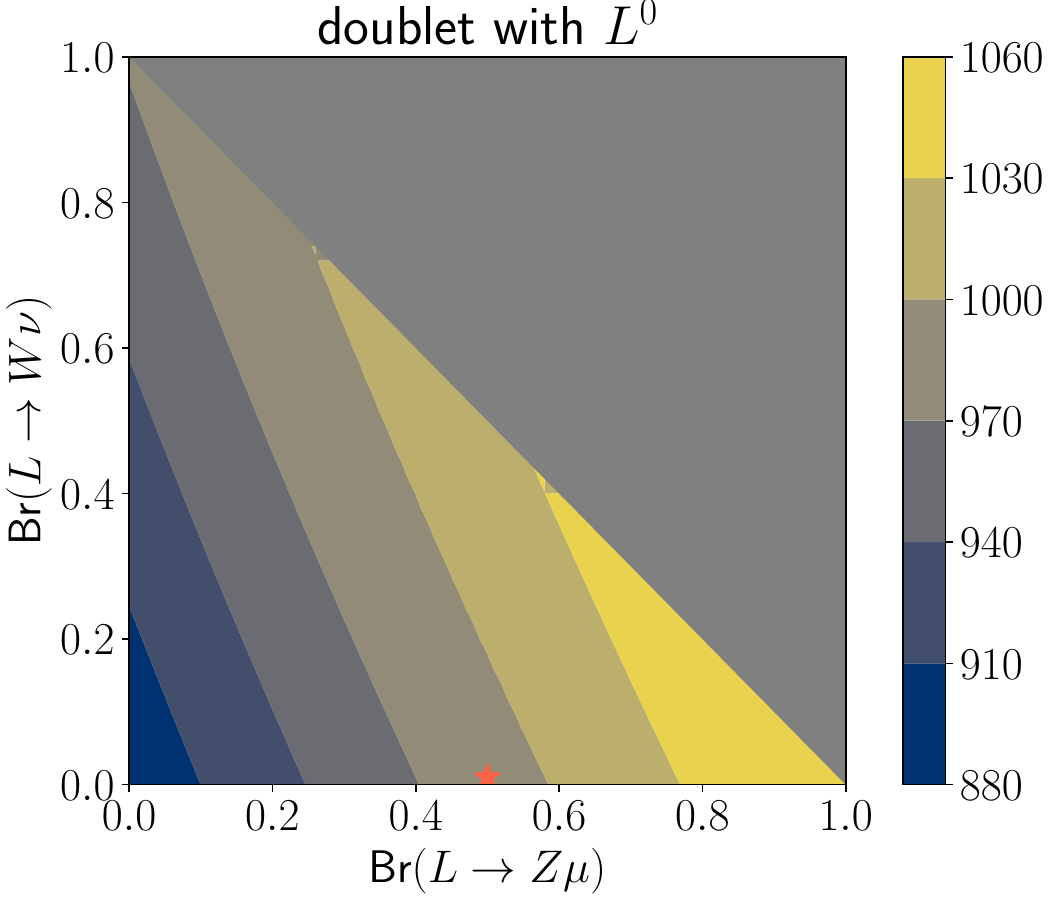}  
\end{minipage}
\caption{\label{fig-HLlims}
The same figures as Fig.~\ref{fig-lims}, but showing the future exclusion limits 
at the HL-LHC, where $\sqrt{s} = 13~\TeV$ and $\Lcal = 3 \mathrm{ab}^{-1}$. 
} 
\end{figure}

Figure~\ref{fig-HLlims} shows the same figure as Fig.~\ref{fig-lims} but with the $3~\mathrm{ab}^{-1}$ data at the future HL-LHC. 
Interestingly, we expect the full running of the HL-LHC can probe a much wider range of the singlet VLL, as shown in the left panel of the figure.
The expected sensitivity on the lower limit of  $m_{VLL}$ is
about 650 GeV for $\br{E}{Z\mu} \sim 1$, 
and it reduces to $100~\GeV$ for $\br{E}{Z\mu} \sim 0.1$. 
It is remarkable that the minimal scenario in Eq.~\eqref{eq-BrLmin} 
can be constrained (discovered) up to about 420 (250) GeV.
From the middle panel, without including the neutral VLL productions, 
the limit is at most 750 GeV  for $\br{L}{Z\mu} \sim 1$ 
and the limit decreases to 100 GeV as $\br{L}{Z\mu}$ decreases 
up to about $0.1$-$0.2$ depending on $\br{L}{W\nu}$.
With the neutral component productions, from the right panel, 
the expected limit is about 1050 GeV for $\br{L}{Z\mu} \sim 1$, 
and decreases to 880 GeV for $\br{L}{Z\mu} \sim 0$.

\section{Summary}
\label{sec-concl}

In this work, we studied the LHC searches for the pair productions of the fourth generation VLLs
which decay to a SM EW boson and a lepton in the second generation. 
We recast the ATLAS analyses searching 
for the triplet lepton in the type-III seesaw~\cite{ATLAS:2020wop,ATLAS:2022yhd}.   
In general, the limits for the VLLs are weaker than those for the triplets 
because the production cross sections are smaller 
and our VLLs are Dirac fermions 
not producing the lepton number violating signals.
In the minimal scenario, shown in Eqs.~\eqref{eq-BrEmin} and~\eqref{eq-BrLmin},
there is no limit beyond about $100~\GeV$ 
on the singlet VLL even with the Run-2 data,  
while the limit is about 750 GeV for the doublet case. 
To obtain the limits for the singlet case using the Run-2 data, 
we should consider more dedicated analysis 
to overcome the small production cross sections, 
e.g. by lowering the threshold for lepton $p_T$'s, 
to be sensitive for the VLL masses within $100$-$200$ GeV, 
and searching for a $3\ell$ resonance decayed from a charged VLL, as done for the Run-1 data\cite{ATLAS:2015qoy}.
For the singlet case, 
it would also be efficient to search for the decays via the SM Higgs boson 
which dominantly decays to a pair of bottom quarks.

We also obtained the expected sensitivities at
 the HL-LHC with $3~\mathrm{ab}^{-1}$ data
and found the future limits would reach to 420 (950) GeV for the singlet (doublet) case. 
It is remarkable that the future experiment will put the limit on the minimal scenario of the singlet VLL.
Note that our analysis covers 
the general cases where the branching fractions 
can be different from those in Eqs.~\eqref{eq-BrEmin} and~\eqref{eq-BrLmin}.
The most significant decay mode is $E^\pm, L^\pm \to Z \mu^\pm$, 
and hence the limits become the most stringent for $\br{E,L}{Z\mu} \sim 1$, 
and are weaker for smaller $\br{E,L}{Z\mu}$.   
Although our results are obtained for the muon-philic VLL scenario, we expect that the sensitivities for the electron-philic VLLs will be basically the same as
those for muon-philic VLLs studied in this paper, because the reconstruction efficiencies for
electrons and muons are similar and all the cuts applied are the same.

The muon-philic VLLs 
are known to be a good candidate to explain the muon $g-2$ anomaly. 
In such a case, the chiral enhancement of the muon $g-2$ by the muon-philic VLLs is correlated to the rate for $h\to \mu^+ \mu^-$,
and the VLLs lighter than about 500 GeV may be excluded~\cite{Dermisek:2013gta,Dermisek:2020cod,Dermisek:2021ajd}.
Our searches can constrain the scenarios of explaining the muon $g-2$ by the chiral enhancement of light VLL correlated with the rate 
for $h\to \mu^+\mu^-$ as well.
The comprehensive study for the VLL explanation of the muon $g-2$ is our future work.

\section*{Acknowledgments} 
The authors thank Radovan Dermisek for useful discussion.
This work was supported by IBS under the project code, IBS-R018-D1.
S.S. acknowledges support from the National Research Foundation of Korea (NRF-2020R1I1A3072747 and NRF-2022R1A4A5030362).


\bibliography{ref}
\bibliographystyle{JHEP} 

\end{document}